\documentclass[aps,showpacs,pre, twocolumn]{revtex4}
\usepackage{amsmath,amssymb,amsthm,amscd,amsfonts}
\usepackage{graphicx}

\begin{document}

\title{Stable dark solitons in $\mathcal{PT}$-symmetric dual-core waveguides}

\author{Yu. V. Bludov$^{1}$, V. V. Konotop$^{2}$, and B. A. Malomed$^{3}$}

\affiliation{$^{1}$Centro de F\'{\i}sica, Universidade do Minho, Campus de Gualtar, Braga
4710-057, Portugal \\
$^{2}$Centro de F\'{\i}sica Te\'{o}rica e Computacional and Departamento de F%
\'{\i}sica, Faculdade de Ci\^encias, Universidade de Lisboa, Avenida
Professor Gama Pinto 2, Lisboa 1649-003, Portugal \\
$^{3}$Department of Physical Electronics, School of Electrical Engineering,
Faculty of Engineering, Tel Aviv University, Tel Aviv 69978, Israel }


\begin{abstract}
We construct dark solitons in the recently introduced model of the nonlinear
dual-core coupler with the mutually balanced gain and loss applied to the
two cores, which is a realization of parity-time symmetry in nonlinear
optics. The main issue is stability of the dark solitons. The modulational
stability of the CW (continuous-wave) background, which supports the dark
solitons, is studied analytically, and the full stability is investigated in
a numerical form, via computation of eigenvalues for modes of small
perturbations. Stability regions are thus identified in the parameter space
of the system, and verified in direct simulations. Collisions between stable
dark solitons are briefly considered too.
\end{abstract}

\maketitle


\section{Introduction}

The concept of the parity-time $\mathcal{PT}$ symmetry was originally
elaborated in the field theory \cite{c:bender2007}, as a generalization of
the canonical conservative systems, which are based on Hermitian
Hamiltonians, for a special case of dissipative systems which include
exactly balanced and spatially separated linear gain and loss. Such systems
are described by non-Hermitian Hamiltonians, whose Hermitian and
anti-Hermitian parts are spatially even and odd, respectively.
A distinctive feature of the non-Hermitian Hamiltonians, which are subject
to the condition of the $\mathcal{PT}$ symmetry, is the fact that, up to a
certain critical value of the strength of their anti-Hermitian (dissipative)
part, the spectrum of such Hamiltonians may remain purely real (physical).
When this occurs a $\mathcal{PT}$-symmetric non-Hermitian Hamiltonian can be
eventually transformed into Hermitian ones by means of similarity
transformations~\cite{43}.

In terms of the quantum theory, $\mathcal{PT}$-symmetric systems are the
settings of theoretical interest. For the realization of the $\mathcal{PT}$
symmetry in real settings, one can make use of the fact that the linear
propagation equation derived for optical beams in the paraxial approximation
has essentially the same form as the Schr\"{o}dinger equation in quantum
mechanics, in one- and two-dimensional (1D and 2D) cases alike. In other
words, the evolution of the wave function of a quantum particle may be
emulated by the transmission of an optical beam, as in both cases the wave
propagation follows the same principles. This fact makes it possible to
simulate many quantum-mechanical phenomena by means of relatively simple
settings which can be realized in classical optics \cite{Longhi}. In this
vein, the realization of $\mathcal{PT}$-symmetric settings in optical
systems, which combine spatially symmetric refractive-index landscapes and
mutually balanced spatially separated gain and loss, was proposed in \cite%
{18} (see also~\cite{18a} for subsequent early development of optical
applications) and experimentally demonstrated in~\cite{c:pt-experim}.

Typically, the models amount to the 1D or 2D linear Schr\"{o}dinger
equations with a complex potential, whose real and imaginary parts are,
respectively, spatially even and odd. Another possibility of the realization
of the $\mathcal{PT}$-symmetric settings in optics, in the form a dual-core
coupler, with the mutually balanced gain and loss applied to the two cores,
was recently proposed in the works~\cite{c:pt-experim,coupler_stat} for
stationary regime of light propagation and in~\cite%
{c:solitons1,c:solitons2,c:solitons3} for the bright optical solitons which
exist when the arms of the coupler obey Kerr nonlinearity. In this last
setting, the solitons are available in the exact analytical form, and their
stability boundary can be found analytically too~\cite%
{c:solitons1,c:solitons3}.



A natural extension of the analysis of the nonlinear $\mathcal{PT}$%
-symmetric systems is to search for stable dark solitons in them, which is
subject of the present work. We notice that the dark solitons in a parabolic
potential with a $\mathcal{PT}$-symmetric non-Hermitian part, where they can
be considered as the nonlinear modes, bifurcating from the first excited
state of the linear $\mathcal{PT}$-symmetric parabolic potential, have
recently been addressed in the literature~\cite{dark}.

An alternative natural setting for the consideration of dark solitons in $%
\mathcal{PT}$-symmetric optical systems is provided by the above-mentioned
dual-core system. As well as a broad class of other solutions, dark solitons
in this system can be easily found in an exact form \cite{c:solitons1}, the
actual problem being the analysis of their stability and interactions. The
model is introduced in Sec. II, and the modulational stability of the CW
(continuous-wave) background, supporting the dark solitons, which is a
necessary condition for their stability, is investigated in an analytical
form in Sec. III. The mathematical framework for the full analysis of the
dark-soliton stability is introduced in Sec. IV, and numerical results,
which can be summarized in the form of stability diagrams for the $\mathcal{%
PT}$-symmetric dark solitons, are reported in Section V. Collisions between
dark solitons are briefly considered in Sec. V too. The paper is concluded
by Sec. VI.

\section{The model}

We start with the system of equations for scaled field variables $q_{1,2}$:
\begin{subequations}
\label{eq:NLS_comp_gen}
\begin{eqnarray}
i\frac{\partial q_{1}}{\partial z} =-\frac{\partial ^{2}q_{1}}{\partial x^{2}%
}+\left( \chi _{1}|q_{1}|^{2}+\chi |q_{2}|^{2}\right) q_{1}+i\gamma
_{1}q_{1}-q_{2},  \label{eq:NLS_comp_gen1} \\
i\frac{\partial q_{2}}{\partial z} =-\frac{\partial ^{2}q_{2}}{\partial x^{2}%
}+\left( \chi |q_{1}|^{2}+\chi _{2}|q_{2}|^{2}\right) q_{2}-i\gamma
_{2}q_{2}-q_{1},  \label{eq:NLS_comp_gen2}
\end{eqnarray}%
Here the linear-coupling constant is scaled to be one, positive coefficients
$\gamma _{1}$ and $\gamma _{2}$ account for the gain and loss, respectively,
in the two cores, while $\chi $ and $\chi _{1,2}$ are real coefficients of
cross-phase modulation (XPM) and self-phase modulation (SPM).

Since the subject of the work is the existence and dynamics of dark
solitons, it is first necessary to address the existence and modulational
stability of the carrier-wave (CW) background, i.e., solutions in the form
of
\end{subequations}
\begin{equation}
q_{1,2}\left( z,x\right) =u_{1,2}\exp \left( -ibz\right) ,  \label{12}
\end{equation}%
with complex amplitudes $u_{1,2}$ and real propagation constant $b$. The
substitution of this into Eqs. (\ref{eq:NLS_comp_gen}) yields
\begin{equation}
|u_{j}|^{2}=\frac{|\gamma _{1}-\gamma _{2}|\sqrt{1-\gamma _{1}\gamma _{2}}}{%
|\gamma _{2}(\chi _{1}-\chi )+\gamma _{1}(\chi -\chi _{2})|}\sqrt{\frac{%
\gamma _{3-j}}{\gamma _{j}}},\quad j=1,2,  \label{background}
\end{equation}%
while the relative phase, $\delta \equiv \arg u_{2}-\arg u_{1}$, is
determined by relation
\begin{equation}
\tan \delta ={\left[ 2\Theta (\gamma _{1}-\gamma _{2})-1\right] }\frac{\sqrt{%
\gamma _{1}\gamma _{2}}}{\sqrt{1-\gamma _{1}\gamma _{2}}},
\end{equation}%
where $\Theta (x)$ is the Heaviside's step function. The propagation
constant of this solution is
\begin{equation}
b=\frac{\cos \delta }{\sqrt{\gamma _{1}\gamma _{2}}}\frac{\gamma
_{1}^{2}\chi _{2}-\gamma _{2}^{2}\chi _{1}}{\gamma _{2}(\chi _{1}-\chi
)+\gamma _{1}(\chi -\chi _{2})}.  \label{b}
\end{equation}%
Note that, according to Eq. (\ref{background}), the CW amplitudes in the two
components are related by $|u_{2}|^{2}/|u_{1}|^{2}=\gamma _{1}/\gamma _{2}$,
which implies the balance between the gain and loss in the CW state.
Further, it follows from Eq. (\ref{background}) that the background
amplitudes have a singularity at $\gamma _{2}/\gamma _{1}=(\chi -\chi
_{2})/(\chi -\chi _{1})\neq 1$, and this solution exists only at $0<\gamma
_{1}\gamma _{2}<1$. This last condition has simple physical explanation: it
requires the gain (dissipation) in an arm to be small enough for being
compensated by the energy flow from the other arm with dissipation (gain),
the flow being limited by the strength of linear coupling (responsible for
the power transfer between the arms) which in our case is normalized to one.

In what follows we concentrate on the case of the $\mathcal{PT}$-symmetry,
with $\gamma _{1}=\gamma _{2}=\gamma $.
Then, it follows from Eq.~(\ref{background}) that the nonzero CW background
may exist only with symmetric SPM coefficients, $\chi _{1}=\chi _{2}$, and
for $\gamma <1$, hence it is convenient to define $\gamma \equiv \sin \delta
$, with $0\leq \delta \leq \pi $, and rewrite Eqs.~(\ref{eq:NLS_comp_gen})
as
\begin{subequations}
\label{eq:NLS_gen}
\begin{eqnarray}
i\frac{\partial q_{1}}{\partial z} &=&-\frac{\partial ^{2}q_{1}}{\partial
x^{2}}+\left( \chi _{1}|q_{1}|^{2}+\chi |q_{2}|^{2}\right) q_{1}+i\sin
(\delta )q_{1}-q_{2},  \notag  \label{eq:NLS_gen2} \\
&& \\
i\frac{\partial q_{2}}{\partial z} &=&-\frac{\partial ^{2}q_{2}}{\partial
x^{2}}+\left( \chi |q_{1}|^{2}+\chi _{1}|q_{2}|^{2}\right) q_{2}-i\sin
(\delta )q_{2}-q_{1}.  \notag \\
&&
\end{eqnarray}

In comparison with the model of the $\mathcal{PT}$-symmetric dual-core
fiber, which was introduced in \cite{c:solitons1,c:solitons2}, Eqs. (\ref%
{eq:NLS_gen}) include the XPM terms, which implies a non-negligible overlap
between transverse modes supported by the two cores. Recently, it was
demonstrated that, in comparison with the well-known results for the
SPM-nonlinear dual-core system with the purely linear coupling \cite%
{c:wabnitz}, the addition of the XPM terms essentially affects the
symmetry-breaking transformations of bright solitons~\cite{bright} and
patterns in the form of domain walls~\cite{DW} in the conservative nonlinear
coupler, whose model amounts to Eqs.~(\ref{eq:NLS_gen}) with $\delta =0$.

\section{Modulational stability of the CW background}

CW solutions of Eqs.~(\ref{eq:NLS_gen}) with equal amplitudes follow from
expressions (\ref{background}):
\end{subequations}
\begin{equation}
q_{j}^{(0)}=\rho \exp\left[i(-1)^j(\delta /2)-ibz\right],\qquad b=\rho
^{2}(\chi _{1}+\chi )-\cos \delta .  \label{eq:pl-wave}
\end{equation}%
Here $j=1,2$, and components have phase mismatch $\delta$ imposed by the
gain-loss coefficient.

To analyze the modulational stability of the CW (\ref{eq:pl-wave}), we use
the standard ansatz with arbitrary real perturbation wavenumber $k$, the
corresponding eigenvalue, $\beta $, and infinitesimal perturbation
amplitudes, $\eta _{j},\,\nu _{j}$:%
\begin{equation}
q_{j}=\rho \left[ e^{i(-1)^{j}\delta /2}+\eta _{j}e^{-i(\beta z-kx)}+\bar{\nu%
}_{j}e^{i(\bar{\beta}z-kx)}\right] e^{-ibz}.
\end{equation}%
Then, two branches $\beta =\beta _{1,2}(k)$ of the linear excitations are
readily found as 
\begin{eqnarray}  \label{omega2}
&&\beta _{1}(k)\equiv \pm k \sqrt{ k^{2}+2\rho^2(\chi_1+\chi)},
\label{omega1} \\
&&\beta _{2}(k)\equiv \pm \sqrt{\left[ k^{2}+2\cos \delta \right] \left[%
k^{2}+2\cos\delta +2\rho ^{2}(\chi _{1}-\chi) \right] }.  \notag \\
\end{eqnarray}

From (\ref{omega1}) it follows that, for the stability of the background,
one has to require
\begin{equation}
\chi _{1}+\chi \geq 0,  \label{chi}
\end{equation}%
the constraint which is also necessary for the modulational stability of the
CW background in the conservative system ($\delta =0$), and which is imposed
in what follows. Equation (\ref{omega2}) gives rise to two other conditions
for the modulational stability,
\begin{equation}
\cos \delta \geq 0,\quad \mathrm{i.e.,}\quad 0\leq \delta \leq \pi /2;
\label{cos}
\end{equation}
\begin{equation}
(\chi _{1}-\chi )\rho ^{2}+\cos \delta >0.  \label{third_cond}
\end{equation}%
\begin{figure}[h]
\begin{center}
\includegraphics*[width=8.5cm]{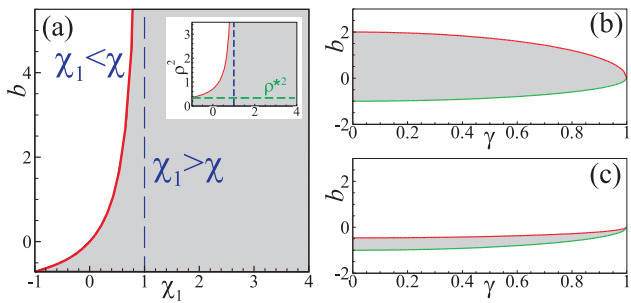}
\end{center}
\caption{Domains of the modulational stability (dashed) and instability
(white) in the $(b,\protect\chi _{1})$ plane for fixed $\protect\gamma %
\equiv \sin \protect\delta =0.7$ (a), and in the $(b,\protect\gamma )$ plane
for fixed $\protect\chi _{1}=0.5$ (b), or $\protect\chi _{1}=-0.3$ (c). In
all the panels $\protect\chi =1$. In panel (a) the left edge corresponds to
the limit form $\protect\chi _{1}+\protect\chi =0$ of condition (\protect\ref%
{chi}), while the bottom edge is given by $b=-\cos (\protect\delta )$. The
inset in panel (a) presents the stability domain in the $(\protect\rho ^{2},%
\protect\chi _{1})$ plane.}
\label{fig:mi-lim}
\end{figure}

In the case of self-focusing SPM, $\chi <0$, the stability domain is
determined by Eqs. (\ref{chi}) and (\ref{cos}) [if these two conditions are
met, Eq. (\ref{third_cond}) is satisfied automatically]. This, in
particular, means that the stability of the CW does not depend on its
amplitude $\rho $, being determined solely by the interplay between the SPM
and XPM coefficients.

The situation is qualitatively different for the defocusing SPM, $\chi >0$.
Now, one can distinguish the two distinct cases. First, if $\chi _{1}>\chi $
[this domain is located to the right from the vertical dashed line in Fig.~%
\ref{fig:mi-lim}(a)], then Eq. (\ref{third_cond}) is reduced to Eq. (\ref%
{cos}), thus giving nothing new, the background being stable at any
amplitude $\rho ^{2}$. If, however, $\chi _{1}<\chi $ [in Fig.~\ref%
{fig:mi-lim}(a), this is the domain to the left from the vertical dashed
line], then, for the stability of the background one needs $\rho ^{2}<\rho
_{\max}^{2}= \cos \delta/(\chi -\chi _{1})$ or, equivalently, $b<2\chi
_{1}\cos \delta /(\chi -\chi _{1})$.

In this situation (i.e., when $\chi _{1}<\chi $) the increase of the
gain-loss coefficient, (i.e. of $\delta $), results in narrowing the
modulational stability domain, which collapses at $\delta=\pi/2$ ($\gamma =1$%
) as shown in Figs.~\ref{fig:mi-lim}(b,c). The limit case of $\chi_1=-\chi$
deserves special consideration, since in this case system (\ref{eq:NLS_gen})
becomes effectively linear for equal amplitudes of components $%
|q_1|=|q_2|=\rho$. As a result, propagation constant $b=-\cos\delta$ does
not depend on $\rho$. Here CW is stable when $\rho^2<{\rho^{\ast}}^{2}=
(\cos \delta)/(2\chi)$. In the same time, value ${\rho^{\ast}}$ defines a
global stability threshold: if $\rho <\rho^{\ast}$, the CW background is
stable at any value of $\chi _{1}>-\chi$ [see the inset in Fig.~\ref%
{fig:mi-lim}(a)].

\section{Stationary dark solitons and their linear stability}

Turning to the study of the dark-soliton solutions, we focus on the
situation when both components have the same intensity profile, i.e.,
\begin{equation}
q_{j}(x,z)=u(x,z)e^{i(-1)^{j}\delta /2}~~(j=1,2),  \label{eq:rel}
\end{equation}%
and thus reduce Eqs. (\ref{eq:NLS_gen}) to the standard nonlinear Schr\"{o}%
dinger equation,
\begin{equation}
i\frac{\partial u}{\partial z}=-\frac{\partial ^{2}u}{\partial x^{2}}+(\chi
_{1}+\chi )|u|^{2}u-\cos (\delta )u,  \label{eq:NLS}
\end{equation}%
whose dark soliton solution is commonly known~\cite{Tsuzuki}:
\begin{equation}
u_{s}(x,z)=\frac{iv-w\tanh \left( w(x-vz)/2\right) }{\sqrt{2(\chi _{1}+\chi )%
}}e^{-ibz}.
\end{equation}%
Here $b$ is given by Eq.~(\ref{eq:pl-wave}), and real parameters $v$ and $w$%
, which determine the \textquotedblleft velocity" (in fact, the spatial
tilt) and the depth of the soliton, are linked by relation ${w}%
^{2}+v^{2}=2(\chi _{1}+\chi )\rho ^{2}$.


Below we focus on the fundamental dark soliton with zero velocity $v=0$
(alias the \textit{black soliton}), $u_{s}(x,z)=u_{0}(x)e^{-ibz}$ where
\begin{equation}
u_{0}(x)=\rho \tanh \left( \rho \sqrt{\frac{\chi _{1}+\chi }{2}}x\right) .
\label{eq:ds}
\end{equation}

To address its stability, we first notice that the CW background must be
modulationally stable, hence the parameters to be considered are limited by
constraints (\ref{chi})-(\ref{third_cond}). Further, to study the linear
stability of the entire dark soliton (\ref{eq:ds}), we adopt the
perturbation solution as
\begin{equation}
q_{j}(x,z)=\left[ u_{0}(x)+u_{j}^{\prime }(x,z)+iu_{j}^{\prime \prime }(x,z)%
\right] e^{i(-1)^{j}\delta /2-ibz},  \label{eq:eig-stab-ans}
\end{equation}%
with infinitesimal perturbation amplitudes $u_{1,2}^{\prime }(x,z)$ and $%
u_{1,2}^{\prime \prime }(x,z)|$. Then, substituting expressions (\ref%
{eq:eig-stab-ans}) into Eq.~(\ref{eq:NLS}), we end up with the eigenvalue
problem:
\begin{equation}
\frac{\partial \mathbf{u}}{\partial z}=\mathcal{L}\mathbf{u},\qquad \mathbf{u%
}=\mathrm{col}\left\{ \,u_{1}^{\prime },u_{1}^{\prime \prime },u_{2}^{\prime
},u_{2}^{\prime \prime }\right\} ,
\end{equation}%
with operators
\begin{eqnarray}
\mathcal{L} &=&\left(
\begin{array}{cccc}
\sin \delta & L_{-} & -\sin \delta & -\cos \delta \\
-L_{+} & \sin \delta & -L & -\sin \delta \\
\sin \delta & -\cos \delta & -\sin \delta & L_{-} \\
-L & \sin \delta & -L_{+} & -\sin \delta%
\end{array}%
\right) , \\
L_{\pm } &\equiv &-\frac{\partial ^{2}}{\partial x^{2}}-b+[(2\pm 1)\chi
_{1}+\chi ]u_{0}^{2}, \\
L &\equiv &2\chi u_{0}^{2}-\cos \delta .
\end{eqnarray}

Let us now prove that the stability analysis can be reduced to two separate
problems,
\begin{equation}
L_{j}\psi =\Lambda _{j}\psi \quad (j=1,2),  \label{stabil_reduced}
\end{equation}%
where the operators are
\begin{subequations}
\label{L12}
\begin{eqnarray}
L_{1} &\equiv &(L_{+}-L)(L_{-}+\cos \delta ),  \label{L2} \\
L_{2} &\equiv &(L_{-}-\cos \delta )(L_{+}+L)
\end{eqnarray}%
such that Im$\Lambda_{1,2} =0$ and Re$\Lambda_{1,2} >0$ constitute necessary
and sufficient conditions for the linear stability of the soliton.

To this end, we notice that SU(4) rotation
\end{subequations}
\begin{equation}
P=\frac{1}{\sqrt{2}}\left(
\begin{array}{rrrr}
0 & -1 & 0 & 1 \\
1 & 0 & 1 & 0 \\
0 & -1 & 0 & -1 \\
-1 & 0 & 1 & 0%
\end{array}%
\right)
\end{equation}%
provides for a unitary transformation, $\mathcal{L}_{0}=P\mathcal{L}P^{-1}$,
with
\begin{equation}
\mathcal{L}_{0}=\left(
\begin{array}{cccc}
0 & 0 & 0 & L-L_{+} \\
0 & 0 & \cos \delta -L_{-} & -2\sin \delta \\
2\sin \delta & L_{+}+L & 0 & 0 \\
L_{-}+\cos \delta & 0 & 0 & 0%
\end{array}%
\right) .
\end{equation}

Since the eigenvalues of $\mathcal{L}$ and $\mathcal{L}_{0}$ coincide, we
can consider the spectrum of the latter linear operator. Taking into account
that both $\mathcal{L}$ and $\mathcal{L}_{0}$ are built of real
coefficients, solutions can be looked for in the form of $u_{1,2}^{\prime
},u_{1,2}^{\prime \prime }\sim \exp (i\lambda z)$. Moreover, if $\lambda $
is an eigenvalue, then $\bar{\lambda}$ is an eigenvalue as well (the overbar
stands for the complex conjugate). In other words, the absence of an
imaginary part of $\lambda $, which is equivalent to the condition that $%
\Lambda =\lambda ^{2}$ is real and positive, is a necessary and sufficient
condition for the absence of the instability.

As the next step, we consider the eigenvalue problem, $-\mathcal{L}%
_{0}^{2}\Psi =\Lambda \Psi $, where $\Psi \equiv P\mathbf{u}$, and we make
use of the block structure of $\mathcal{L}_{0}^{2}$:
\begin{equation}
-\mathcal{L}_{0}^{2}=\left(
\begin{array}{cccc}
L_{1} & 0 & 0 & 0 \\
4\sin (\delta )L_{-} & L_{2} & 0 & 0 \\
0 & 0 & L_{2} & 4\sin (\delta )L_{+} \\
0 & 0 & 0 & L_{1}%
\end{array}%
\right) ,
\end{equation}%
where $L_{1,2}$ were introduced in Eqs. (\ref{L12}). Now, a straightforward
consideration demonstrates that $\Lambda $ must coincide with either $%
\Lambda _{1}$ or $\Lambda _{2}$. Thus, the study of the stability of the
dark solitons is reduced to eigenvalue problems (\ref{stabil_reduced}).

Now, we notice that
\begin{subequations}
\label{L2_new}
\begin{eqnarray}
L_{-}-\cos \delta &=&-\frac{\partial ^{2}}{\partial x^{2}}-(\chi _{1}+\chi
)(\rho ^{2}-u_{0}^{2}), \\
L_{+}+L &=&L_{0}+2(\chi _{1}+\chi )u_{0}^{2}.
\end{eqnarray}%
Therefore, taking into account Eq. (\ref{chi}), we conclude that the
eigenvalue problem for operator $L_{2}$ is nothing but the standard
stability problem for the black soliton in the defocusing medium, with the
effective nonlinearity $\chi _{1}+\chi $. This problem is very well studied~%
\cite{Barash_stabil,Chen_stabili}. In particular, it is known that $L_{+}+L$
is positive definite and $L_{-}-\cos \delta $ has only one negative
eigenvalue and one zero eigenvalue~\cite{Barash_stabil}. Moreover, it is
known~too \cite{Chen_stabili} that the minimal eigenvalue of $L_{2}$ is
positive. Thus, the eigenvalue problem for $L_{2}$ does not give
instability, and our subsequent analysis is performed below for operator $%
L_{1}$, which may give rise to instability.

\section{Numerical results}

The results of the numerical analysis of the linear stability are depicted
in Fig.~\ref{fig:ds-lim}. For the defocusing XPM, 
in the subdomain $-\chi \leq \chi _{1}\leq \chi _{1}^{\ast }$ (where $\chi
_{1}^{\ast }$ is the critical value, denoted in Fig.~\ref{fig:ds-lim}(a) by
the blue vertical line), the dark-soliton stability region coincides with
that for the CW background, which is $-\cos \delta \leq b\leq 2\chi
_{1}\left( \cos \delta \right) /(\chi -\chi _{1})$ [or, equivalently, $\rho
^{2}\leq \left( \cos \delta \right) /(\chi -\chi _{1})$] for $\chi_1<\chi$
and $b\geq-\cos \delta$ for $\chi\leq \chi_1\leq \chi _{1}^{\ast }$, see Fig.%
\ref{fig:mi-lim}(a). At the same time, in subdomain $\chi _{1}\geq \chi
_{1}^{\ast }$ a dark-soliton's instability ``wedge" is present: as seen from
Fig. \ref{fig:ds-lim}(a), the dark soliton is stable when $-\cos \delta \leq
b\leq b_{1}$ or $b_{2}\leq b<\infty $. The value of the propagation
constant, $b_{1}$, at the lower edge of the ``wedge" [the green line in Fig. %
\ref{fig:ds-lim}(a)] is almost independent of SPM coefficient $\chi _{1}$
(except for a small region in a vicinity of the critical value $\chi
_{1}^{\ast }$), while the upper edge, $b=b_{2}$ [the red line in Fig. \ref%
{fig:ds-lim}(a)], is a quasi-linear function of $\chi _{1}$. With the
increase of $\gamma =\sin \delta $ the dark-soliton instability ``wedge"
gradually shrinks [see Fig. \ref{fig:ds-lim}(b)], disappearing at $\gamma =1$%
. For the focusing or zero XPM, with $\chi =-1$ or $\chi =0$, the dark
soliton is stable at $-\cos \delta \leq b\leq b_{1}$, and unstable at $%
b>b_{1}$, see Fig. \ref{fig:ds-lim}(c). It is relevant to note that $b_{1}$
does not depend on $\chi _{1}$, and almost coincides with $b_{1}$
corresponding to the defocusing XPM [cf. Figs. \ref{fig:ds-lim}(b) and \ref%
{fig:ds-lim}(c)], while $b_{2}$ coincides with the vertical line, $\chi
_{1}=-\chi $.

\begin{figure}[t]
\begin{center}
\includegraphics*[width=8.5cm]{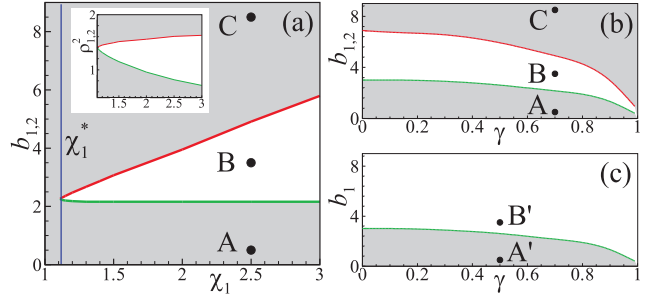}
\end{center}
\caption{Regions of the stability and instability of the dark soliton
(domains covered by dashed patterns and white ones, respectively) in the $(b,%
\protect\chi _{1})$ plane for fixed $\protect\gamma =0.7$ and $\protect\chi %
=1$ (a), and in the $(b,\protect\gamma )$-plane for fixed $\protect\chi %
_{1}=2.5$, $\protect\chi =1$ (b), or $\protect\chi =-1$ (c). In panel (a)
inset presents the same regions as the main panel, but in the $(\protect\rho %
^{2},\protect\chi _{1})$ plane;$\protect\chi _{1}^{\ast }$ coincides
with the left edge of the inset.} \label{fig:ds-lim}
\end{figure}

The linear stability analysis was completed by the direct simulations of
Eqs.~(\ref{eq:NLS_gen}). Typical examples of the perturbed evolution of
stable and unstable dark solitons are presented in Fig.~\ref{fig:ds-chip}.
The predicted stability of the dark soliton below the lower edge of the
instability ``wedge", i.e., at $b<b_{1}$, is confirmed by the simulations.
For the dark-soliton parameters corresponding to point A in Figs.~\ref%
{fig:ds-lim}(a) and \ref{fig:ds-lim}(b), the evolution of fields component $%
q_{1,2}(x,z)$ is shown in Fig. \ref{fig:ds-chip}(a). Similarly, the
evolution of the dark soliton with parameters corresponding to point B, as
well as the stability of the dark soliton with the parameters corresponding
to point C in Figs. \ref{fig:ds-lim}(a) and \ref{fig:ds-lim}(b), are
demonstrated in Figs.~\ref{fig:ds-chip}(b) and \ref{fig:ds-chip}(c),
respectively. 
\begin{figure}[t]
\begin{center}
\includegraphics*[width=8.5cm]{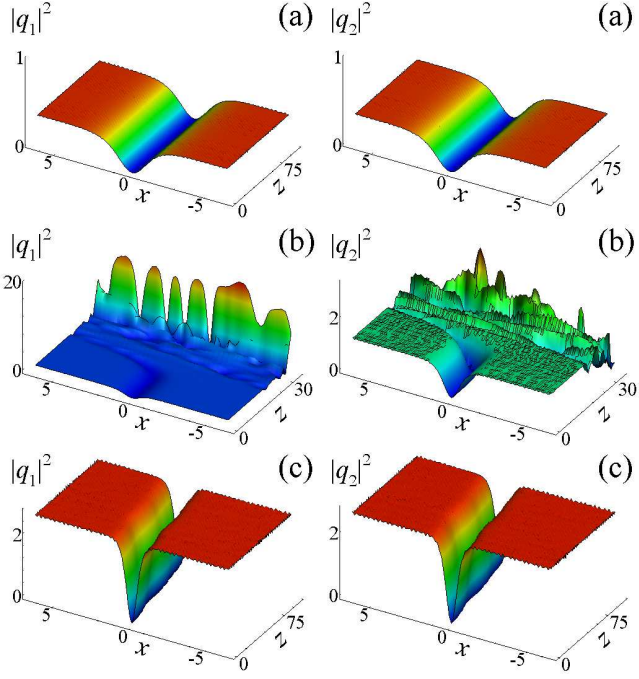}
\end{center}
\caption{The evolution of field components $|q_{1}(x,z)|^{2}$ (left column)
and $|q_{2}(x,z)|^{2}$ (right column) of dark soliton (\protect\ref{eq:ds})
with $\protect\gamma =0.7$, $\protect\chi _{1}=2.5$, $\protect\chi =1$, $%
b=0.5$, $\protect\rho ^{2}\approx 0.347$ (a), $b=3.5$, $\protect\rho %
^{2}\approx 1.2$ (b), or $b=8.5$, $\protect\rho ^{2}\approx 2.633$ (c).
Panels (a), (b) and (c) correspond to points A, B, and C in Fig. \protect\ref%
{fig:ds-lim}, respectively. }\label{fig:ds-chip}
\end{figure}

For the focusing XPM, the stability and instability of the dark solitons
[points A$^{\prime }$ and B$^{\prime }$ in Fig. \ref{fig:ds-lim}(c)] is also
confirmed by direct simulations, see Figs.~\ref{fig:ds-chim}(a) and \ref%
{fig:ds-chim}(b), respectively.
\begin{figure}[t]
\begin{center}
\includegraphics*[width=8.5cm]{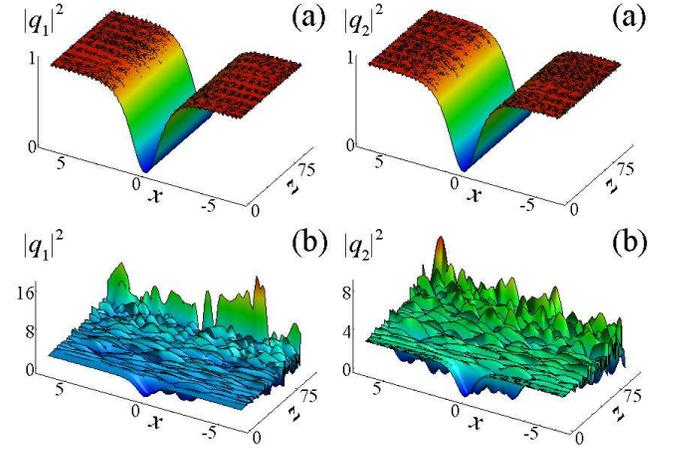}
\end{center}
\caption{The same as in Fig.\protect\ref{fig:ds-chip}, but for $\protect%
\gamma =0.5$, $\protect\chi _{1}=2.5$, $\protect\chi =-1$, $b=0.5$, $\protect%
\rho ^{2}\approx 0.910$ (a), or $b=3.5$, $\protect\rho ^{2}\approx
2.910$ (b). Panels (a) and (b) correspond to points A$^{\prime }$
and B$^{\prime }$ in Fig. \protect\ref{fig:ds-lim}, respectively. }
\label{fig:ds-chim}
\end{figure}

The robustness of the dark solitons can be also be tested in interactions of
two such solitons (kink-antikink pairs). Thus, in this case we use the
initial condition at $z=0$ in the form of
\end{subequations}
\begin{eqnarray}
u(x) &=&\rho \left[ \tanh \left\{ \rho \sqrt{\frac{\chi _{1}+\chi }{2}}%
\left( x+\frac{\ell}{2}\right) \right\} -\right.  \notag \\
&&\left. \tanh \left\{ \rho \sqrt{\frac{\chi _{1}+\chi }{2}}\left( x-\frac{%
\ell }{2}\right) \right\} -1\right] ,  \label{eq:ds-two}
\end{eqnarray}%
where $\ell$ is the spatial separation between the two dark solitons.
\begin{figure}[t]
\begin{center}
\includegraphics*[width=8.5cm]{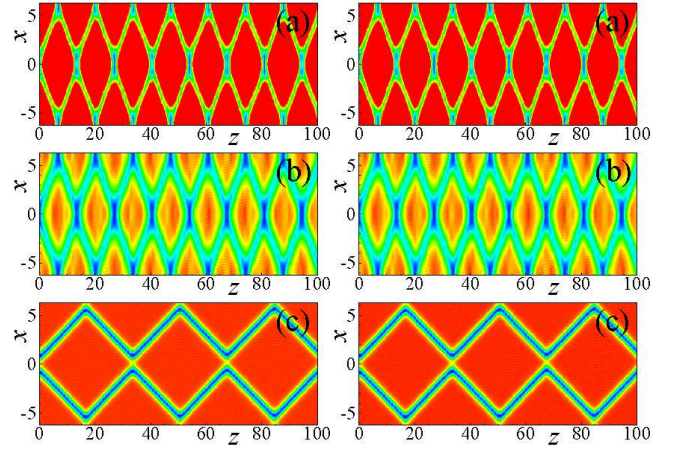}
\end{center}
\caption{The evolution of field components $|q_{1}(x,z)|^{2}$ (left column)
and $|q_{2}(x,z)|^{2}$ (right column) of dark-soliton pair (\protect\ref%
{eq:ds-two}) with $\protect\chi _{1}=2.5$, $\ell =\protect\pi /2$ and $%
\protect\chi =-1$, $\protect\gamma =0.5$, $b=0.5$, $\protect\rho ^{2}\approx
0.910$ (a), $\protect\chi =1$, $\protect\gamma =0.7$, $b=0.5$, $\protect\rho %
^{2}\approx 0.347$ (b), or $\protect\chi =1$, $\protect\gamma =0.7$, $b=8.5$%
, $\protect\rho ^{2}\approx 2.633$ (c). Panels (a), (b) and (c) correspond
to points A$^{\prime }$, A, and C in Fig.~\protect\ref{fig:ds-lim},
respectively. }
\label{fig:ds-two}
\end{figure}
As can be seen from Fig. \ref{fig:ds-two}, in the $\mathcal{PT}$-symmetric
system the two dark soliton \emph{always} [for both the focusing -- Fig. \ref%
{fig:ds-two}(a) and defocusing -- Figs. \ref{fig:ds-two}(b), \ref{fig:ds-two}%
(c) signs of the XPM] repel each other and start motion in opposite
directions without self-destruction. The repulsion from the boundaries of
the $x$-domain in Fig.~\ref{fig:ds-two} happens due to the implied periodic
boundary conditions, and is equivalent to the repulsion between the dark
soliton. As can be seen from the comparison of Figs.~\ref{fig:ds-two}(b) and %
\ref{fig:ds-two}(c), the increase of $b$ (while separation $\ell$ between
the dark solitons is kept unchanged) results in reduction of the repulsion
between the solitons, and, consequently, decrease of the solitons'
``velocities". The reason for this phenomenon is that larger $b$ corresponds
to a smaller soliton width, and, as a result, a larger ratio of separation $%
\ell$ to the soliton's width. 
\begin{figure}[t]
\begin{center}
\includegraphics*[width=8.5cm]{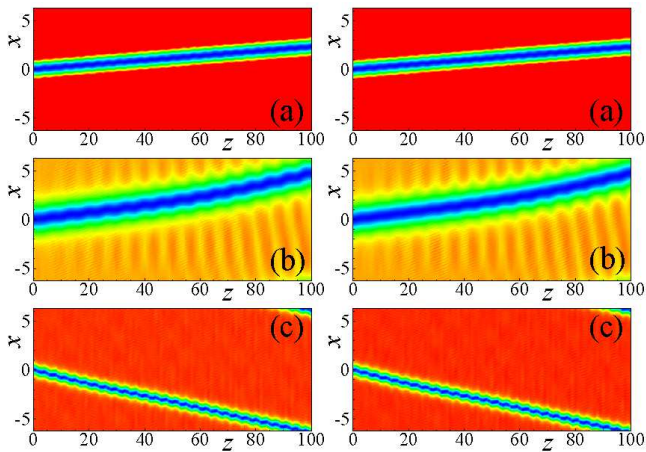}
\end{center}
\caption{The evolution of field components $|q_{1}(x,z)|^{2}$ (left column)
and $|q_{2}(x,z)|^{2}$ (right column) of the dark soliton with initially
separated components (\protect\ref{eq:ds-sep}), for $D=\protect\pi /50$.
Other parameters are the same as in Fig. \protect\ref{fig:ds-two}.}
\label{fig:ds-sep}
\end{figure}

Another possibility to set dark soliton in motion is to separate components
in the initial condition, i.e., take
\begin{equation}
q_{j}(x,0)=u(x+(-1)^{j}D/2)e^{i(-1)^{j}\delta /2},  \label{eq:ds-sep}
\end{equation}%
where $u(x)$ is borrowed from Eq.(\ref{eq:ds}), and $D$ is the initially
imposed separation between the components. The results of the corresponding
simulations are represented in Fig. \ref{fig:ds-sep}. Here, for the
defocusing XPM the ``velocity" of the dark soliton does not strongly depend
upon propagation constant $b$ [cf. Figs. \ref{eq:ds-sep}(b) and \ref%
{fig:ds-sep}(c)]. The situation is completely different from the previous
case [cf. Figs. \ref{fig:ds-two}(b) and \ref{fig:ds-two}(c)]. It should be
noted that, in both Figs. \ref{fig:ds-two} and \ref{fig:ds-sep}, the
simulations were run for dark solitons with the propagation constants far
enough from the stability margins $b_{1,2}$ [points A$^{\prime }$, A and C
in Fig.~\ref{fig:ds-lim}]. In the opposite situation, for the dark solitons
with propagation constants close to stability margins $b_{1,2}$, their
motion may result in destruction, under certain conditions.

\section{Conclusions}

To conclude we have reported the existence of stable vector solitons in the $%
\mathcal{PT}$-symmetric coupled nonlinear Schr\"{o}dinger equation one of
which has gain and another dissipation, whose strengths are equal. The found
solitons have identical amplitude profiles but the phase difference imposed
by the gain-loss coefficients what ensures the balance between gain and
losses. The stability of either backgrounds against which solitons propagate
or of the solitons themselves are modified by dissipation and gain, what was
confirmed by direct numerical simulations of the soliton propagation and
interactions as well by the linear stability analysis.

\section*{Acknowledgements}

VVK acknowledges support of Funda\c{c}\~{a}o para a Ci\^{e}ncia e a
Tecnologia (Portugal) under grants PTDC/FIS/112624/2009 and
PEst-OE/FIS/UI0618/2011.

\end{document}